\documentclass[oldversion,letter]{aa}  
\usepackage{natbib}
\usepackage{graphicx}
\usepackage{txfonts}
\newcommand{\ms}{$\,$M$_\mathrm{\odot}$}

\newcommand{\be}{\begin{equation}}
\newcommand{\ee}{\end{equation}}

\newcommand{\gad}{\ensuremath{\bigtriangledown_\mathrm{ad}}}
\newcommand{\el}[2]{\ensuremath{^{#1}\mathrm{#2}}}

\begin{document}
   \title{Carbon-enhanced metal-poor stars and thermohaline mixing}


   \author{R.~J. Stancliffe
          \inst{1}
          \and
          E. Glebbeek\inst{2}
          \and
          R.~G. Izzard\inst{2}
          \and
          O.~R. Pols\inst{2}
          }

   \offprints{R.~J. Stancliffe}

   \institute{Institute of Astronomy, The Observatories, Madingley Road, Cambridge, CB3 0DS, U.K.\\
              \email{rs@ast.cam.ac.uk}
         \and
             Sterrekundig Instituut Utrecht, Postbus 80000, 3508 TA Utrecht, The Netherlands.\\
             \email{E.Glebbeek@phys.uu.nl, R.G.Izzard@phys.uu.nl, O.R.Pols@phys.uu.nl}
             }

   \date{Received 7 December 2006; Accepted 31 January 2007}

\abstract{One possible scenario for the formation of carbon-enhanced metal-poor stars is the accretion of carbon-rich material from a binary companion which may no longer visible. It is generally assumed that the accreted material remains on the surface of the star and does not mix with the interior until first dredge-up. However, thermohaline mixing should mix the accreted material with the original stellar material as it has a higher mean molecular weight. We investigate the effect that this has on the surface abundances by modelling a binary system of metallicity $Z=10^{-4}$ with a 2\ms\ primary star and a 0.74\ms\ secondary star in an initial orbit of 4000 days. The accretion of material from the wind of the primary leads to the formation of a carbon-rich secondary. We find that the accreted material mixes fairly rapidly throughout 90\% of  the star, with important consequences for the surface composition. Models with thermohaline mixing predict very different surface abundances after first dredge-up compared to canonical models of stellar evolution.}

   \maketitle
%

\section{Introduction}
Carbon-enhanced, metal-poor (CEMP) stars are defined as stars with [C/Fe]$>$+1.0 \citep{2005ARA&A..43..531B}. A possible formation scenario for these stars is the accretion of carbon-rich material from a companion. It is commonly assumed \citep[e.g.][]{2004A&A...416.1117C,2005ApJ...625..833L} that this accreted material sits unmixed on the surface of the star up to the point at which the star undergoes first dredge-up on the red giant branch. At this point, it is thought that the accreted material becomes mixed with part of the interior of the star in the deep convective envelope that develops. However, this picture neglects thermohaline mixing.

Thermohaline mixing is the process that occurs when the mean molecular
weight of the stellar gas increases towards the surface. Although this
situation is dynamically stable as long as the density decreases
outward, a secular instability sets in when a gas element is displaced
downwards and compressed. Such an element must be hotter than the
surrounding gas by virtue of its higher molecular weight and if it
loses heat to its surroundings its density increases and it will
continue to sink. As a result mixing occurs on a thermal timescale,
until the molecular weight difference has disappeared
\citep{1980A&A....91..175K}. In interacting binary systems, mass transfer (by wind accretion or Roche lobe overflow) may lead to the transfer of material that has been processed by nuclear burning from the primary to the surface of the less evolved secondary. This alters the surface composition of the secondary. The extent to which the surface composition is changed depends on how much the transferred matter is mixed with the pristine stellar material.

We expect thermohaline mixing to occur in a star that has accreted material from a carbon-rich companion as this material will have a greater mean molecular weight than that of the unevolved star. The aim of this work is to determine the effect that thermohaline mixing has on the surface abundances of such an object. In section 2, we describe our model. In section 3, we describe the results and in section 4 these are discussed in the context of observations of CEMP stars.

\section{The model}
We use the rapid synthetic binary code of \citet{2006A&A...460..565I} to model the evolution of a binary system of metallicity $Z=10^{-4}$ consisting of a 2\ms\ primary with a 0.74\ms\ companion in a circular orbit with an initial period of 4000 days. The secondary mass is chosen as being representative of a companion with a mass close to the turn-off mass of the halo. The primary mass was chosen as an example of an asymptotic giant branch (AGB) star that becomes very C-rich. During the thermally pulsing asymptotic giant branch (TP-AGB) the primary undergoes third dredge-up and becomes carbon rich while the envelope is eroded by the stellar wind. Based on the Bondi-Hoyle \citep{1944MNRAS.104..273B} prescription for wind accretion {(as used in the synthetic code, see \citealt{2006A&A...460..565I} and references therein)}, we find that the secondary accretes 0.09\ms\ of material from the wind of the primary and we record the average composition of this material (see Table~\ref{tab:compos}). { The total mass of material accreted and the average composition of this material are then used in creating a detailed model of the secondary.}

We model the secondary using a version of the stellar evolution code originally developed by \citet{1971MNRAS.151..351E} and updated by many authors \citep[e.g.][]{1995MNRAS.274..964P}. Our version includes the nucleosynthesis subroutines of \citet{2005MNRAS.360..375S} and \citet{stancliffe05}. { The code uses the opacity routines of \citet{2004MNRAS.348..201E}, which employ interpolation in the OPAL tables \citep{1996ApJ...464..943I} and which account for the variation in opacity as the C and O content of the material varies}. We create a 0.74\ms\ model and evolve it from the pre-main sequence to the age of the system predicted by the rapid code.  The initial abundances are chosen to be solar-scaled based on the values of \citet{1989GeCoA..53..197A}, with [Fe/H]\footnote{This is the logarithm of the ratio of the iron abundance to the hydrogen abundance, relative to the solar value, i.e. [Fe/H] = $\log (X_\mathrm{Fe}/X_\mathrm{H}) - \log(X_\mathrm{Fe}/X_\mathrm{H})_\odot$.}=-2.3. A mixing length parameter of $\alpha=2.0$ was used and no convective overshooting was employed.

Thermohaline mixing was included according to the prescription of \citet{1980A&A....91..175K}. They model it as a diffusive process with a diffusion coefficient, $D_\mathrm{th}$, that can be written as
\be
D_\mathrm{th} = {16acT^3H_\mathrm{P}\over (\gad - \bigtriangledown)c_\mathrm{P}\rho\kappa} \left| \mathrm{d}\mu \over \mathrm{d}r \right| {1\over \mu},
\label{eq:dth}
\ee
where $H_\mathrm{P}$ is the pressure scale height, $a$ is the radiation pressure constant, $c$ is the speed of light, $T$ is the temperature, $\gad$ is the adiabatic temperature gradient, $\bigtriangledown$ is the actual temperature gradient,  $c_\mathrm{P}$ is the specific heat at constant pressure, $\rho$ is the density, $\kappa$ is the opacity, $r$ is the radius and $\mu$ is the mean molecular weight of the material. 

Once we have a model of the desired age (specifically $8.6\times10^{8}$\,yrs), we create two model sequences. In both, we accrete 0.09\ms\ of material with the composition given by the rapid code. The accretion rate is set to $10^{-6}$\ms\,yr$^{-1}$, which is approximately the rate at which the secondary would accrete material from the AGB superwind. In one sequence we include the effects of thermohaline mixing; in the other we do not.

\section{Results}

\begin{figure}
\centering
\includegraphics[width=8cm]{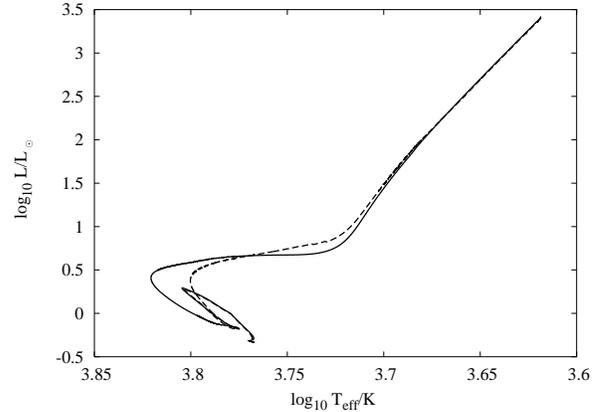}
\caption{Hertzsprung-Russell diagram showing the evolution of the two models. The model sequence with thermohaline mixing is the solid line; the dashed line is the model sequence without thermohaline mixing. The excursion at the beginning of the evolutionary tracks represents the accretion phase.}
\label{fig:HR}
\end{figure}

A Hertzsprung-Russell (HR) diagram showing the evolution of the two models is presented in Figure~\ref{fig:HR}. The model without thermohaline mixing (dashed line) is cooler than the model with thermohaline mixing (solid line). This is because the accreted material is not mixed into the interior of the star. Although the star has a surface convection zone, this is too shallow to cause significant mixing. Because the accreted material has a higher opacity {on account of it being extremely carbon rich}, the surface temperature of the model without thermohaline mixing is lower and the star's position is shifted in the HR diagram relative to that of the model with thermohaline mixing.

\begin{figure}
\centering
\includegraphics[width=8cm]{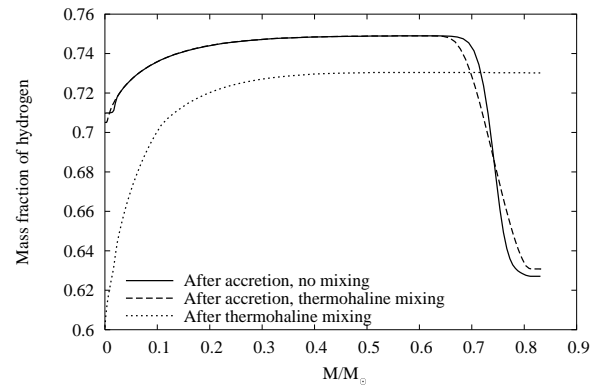}
\caption{Profile of the H abundance as a function of mass for the secondary. The abundance profile immediately after accretion is shown as solid line. The model with thermohaline mixing is the dashed line. The dotted line shows the hydrogen abundance once thermohaline mixing has ceased. This model is taken at about $10^9$ years after the initial model.}
\label{fig:profiles}
\end{figure}

Figure~\ref{fig:profiles} shows the hydrogen abundance profile of the
star just after accretion for both models. Before accretion the
surface material has a mean molecular weight of $\mu = 0.593$ whilst
the accreted material has a mean molecular weight of $\mu=0.657$
(mainly caused by its higher helium abundance). Thermohaline mixing is
strong enough to affect the profile during the accretion phase so the
model with thermohaline mixing (dashed line) has a slightly shallower
hydrogen abundance profile near the surface after accretion than the
unmixed model (solid line). Figure~\ref{fig:profiles} also shows the
H-abundance profile when thermohaline mixing has ceased (dotted
line). Note that by the time this state is reached, significant
H-burning has occurred in the core.

We find that thermohaline mixing results in the star being mixed down to
within 0.1\ms\ of the centre. This represents almost 90\% of the star.
Only the very central material escapes being mixed as it is here that
material has experienced sufficient nuclear burning to raise its mean
molecular weight. The mixing is rapid and the star reaches its equilibrium
configuration within about $10^9$ years after accretion. This is about 
one-tenth of its main-sequence lifetime.

The ratio of surface abundances to iron (relative to solar) of the model
sequences at the end of core hydrogen burning are shown in
Table~\ref{tab:compos}. Thermohaline mixing has a significant effect on
the surface abundances. In the model with thermohaline mixing all heavy
elements are diluted compared to the unmixed model, but the mixing does not 
dilute all the abundances by the same amount -- the degree of dilution
depends on how different the abundance of the original material is from that of the 
accreted material. For instance, the accreted material is extremely
carbon-rich and when it is mixed with the pristine stellar material a large change in
the abundance results: [C/Fe] is reduced from $3.25$ in the model without
thermohaline mixing to $2.41$ in the mixed model. Similarly [Na/Fe] and [Mg/Fe]
are reduced by about $0.8$ dex. For other elements the change is less severe.

\begin{table*}
\centering
\begin{tabular}{l|cccccc}
 & [C/Fe] & [N/Fe] & [O/Fe] & [Na/Fe] & [Mg/Fe] & [Al/Fe] \\
 \hline
Accreted material & 3.26 & 0.40 & 0.99 & 2.11 & 1.68 & 1.02 \\
\hline
With thermohaline mixing: \\
End of core H-burning & 2.41 & 0.10 & 0.35 & 1.27 & 0.88 & 0.36 \\
After first dredge-up & 2.33 & 1.91 & 0.34 & 1.39 & 0.87 & 0.36 \\
\hline
Without thermohaline mixing: \\
End of core H-burning & 3.25 & 0.39 & 0.98 & 2.10 & 1.67 & 1.01 \\
After first dredge-up & 2.47 & 0.15 & 0.37 & 1.33 & 0.93 & 0.39 \\
\hline
\end{tabular}
\caption{Ratio of surface metal abundaces to iron measured relative to solar values. Before accretion, the ratios are zero as we have assumed solar-scaled initial values.}
\label{tab:compos}
\end{table*}

\subsection{Abundances after first dredge-up}
As a low-mass star begins to ascend the red giant branch, the convective
envelope deepens and material from the interior of the star is drawn to
the surface. This is first dredge-up. Material that has experienced
CN-cycling is brought to the surface and hence the surface carbon
abundance drops while the abundance of nitrogen increases. If accreted
material were to remain unmixed on the surface of the star then during
first dredge-up it would become mixed with the interior of the star. It is
for this reason that surveys like that of \citet{2005ApJ...625..833L}
select dwarf stars, rather than giant stars.

\begin{figure*}
\centering
\includegraphics[width=8cm]{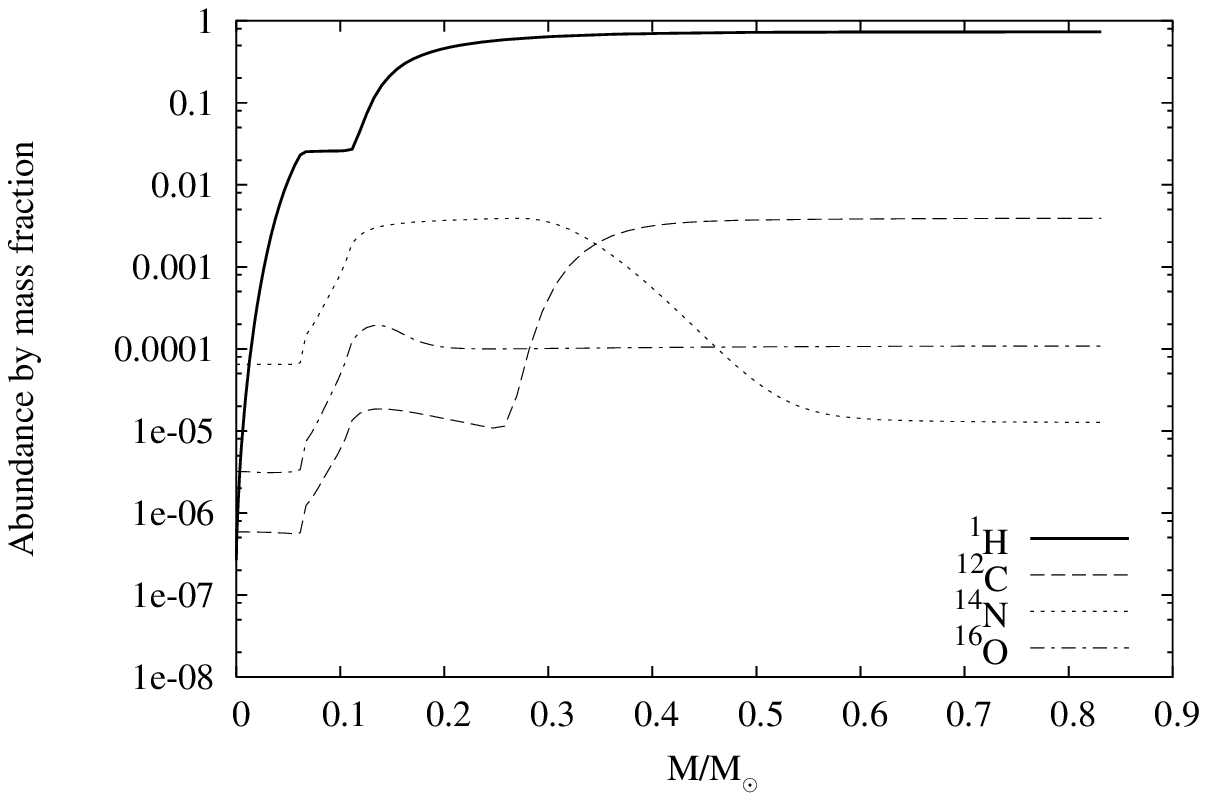}
\includegraphics[width=8cm]{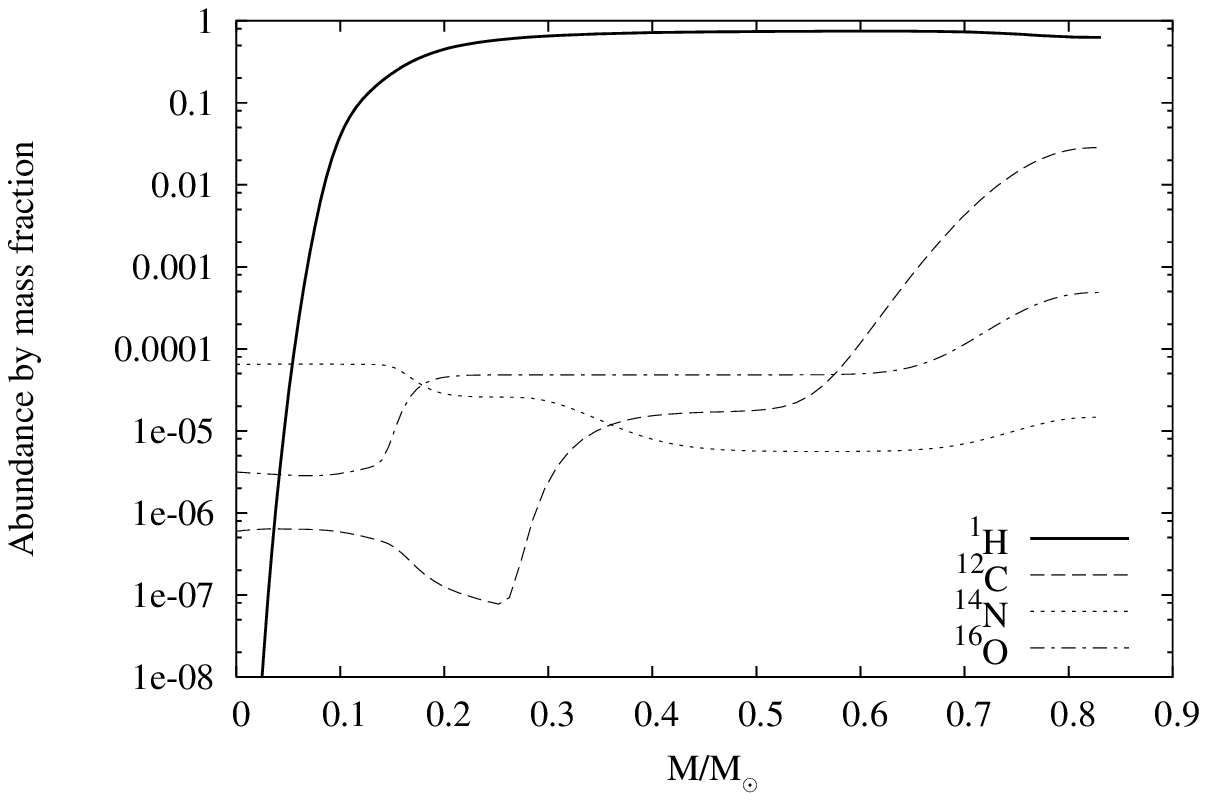}
\caption{Profiles of the {\el{ 1}{ H}, \el{ 12}{ C}, \el{ 14}{ N} and \el{ 16}{ O}} abundances as a fuction of mass at the end of the main-sequence. The left-hand panel is the model with thermohaline mixing and the right-hand panel is the model without thermohaline mixing. The step in the hydrogen profile of the thermohaline mixing model  is due to the existence of a gradient in the abundance of CNO. The burning rate increases rapidly across this region due to the increased abundance in these metals. The models are not quite co-incident in time due to the frequency at which the code outputs detailed models.}
\label{fig:predupprofile}
\end{figure*}

Table~\ref{tab:compos} lists the surface abundances of the two models
after the occurrence of first dredge-up. For the model without thermohaline
mixing the main effect of the dredge-up is not to bring processed material
to the surface, but to dilute the formely accreted material by
mixing it with the rest of the envelope -- the surface abundance of all elements
is decreased. The post-dredge-up abundances resemble those of the model with thermohaline mixing at the end of
the main sequence, although the dilution is less severe because the
dredge-up does not mix as deeply.

Conversely, in the model that does include thermohaline mixing the accreted
material had already been mixed during the main sequence, so the dredge-up
acts like normal, bringing processed material to the surface.
Some of the accreted C-rich material was mixed into the interior where it
reached temperatures high enough for CN-cycling to occur (see the left-hand 
panel of Figure~\ref{fig:predupprofile}). This material is brought to the
surface again by the dredge-up, causing an increase in [N/Fe] and a
decrease in [C/Fe]. The [N/Fe] before dredge-up is
primarily determined by the nitrogen content of the accreted material,
whereas [N/Fe] after dredge-up is primarily determined by the carbon content of
the accreted material.

The thermohaline mixing model shows a slight enhancement in [Na/Fe] for a
similar reason. The accreted material is rich in \el{22}{Ne} (the abundance of 
\el{22}{Ne} by mass fraction in the accreted material is
$2.1\times10^{-3}$). When this material is mixed into the interior a small amount of this
\el{22}{Ne} is converted to \el{23}{Na} by proton capture, even though the reaction
rate is small. Since the \el{23}{Na} abundance is low (its abundance by mass
fraction is around $10^{-7}$ in the original stellar material) even a small amount of 
conversion of \el{22}{Ne} contributes significantly to the abundance of 
\el{23}{Na}.

\section{Comparison to observations}

\begin{figure}
\centering
\includegraphics[width=8cm]{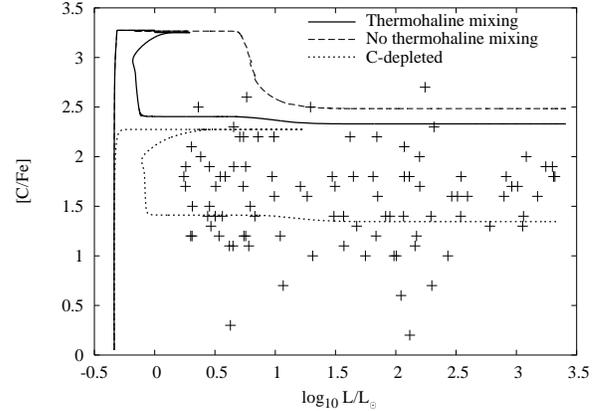}
\caption{Plot of the evolution of the [C/Fe] ratios of the two models as a function of luminosity. The solid line is the model with thermohaline mixing; the dashed line is the model without thermohaline mixing. The dotted line is a model with the [C/Fe] ratio of the accreted material decreased by one dex and thermohaline mixing included. Crosses represent the carbon-rich stars of the HERES sample \citep{2006ApJ...652L..37L}. The decrease in the [C/Fe] ratio of the model with thermohaline mixing at $\log_{10} L\mathrm{/L_{\odot}}\approx0$ takes around $10^9$\,yrs. The excursion to higher luminosity before [C/Fe] drops corresponds to the accretion phase, which is very rapid compared to the rest of the evolution.} 
\label{fig:CFeversusL}
\end{figure}

We now compare our models to observations of CEMP
stars. \citet{2006ApJ...652L..37L} published a large sample of CEMP
stars and their abundances as a function of luminosity. If we assume
that this data set reflects the chemical evolution of these objects,
our models should reproduce the trend seen in these data. 
Figure~\ref{fig:CFeversusL} shows the evolution of the [C/Fe] ratio of
our models against the Lucatello et al. data. The model without
thermohaline mixing (dashed line) displays a large drop in [C/Fe]
which begins at around $\log_{10} L\mathrm{/L_{\odot}}=0.7$ due to the
onset of first dredge-up, when the accreted material is mixed into the
stellar interior as the convective envelope deepens. In the model with
thermohaline mixing (solid line) the accreted material is mixed into
the interior whilst the star is still on the main sequence, at
$\log_{10} L\mathrm{/L_{\odot}}\approx 0$. Thus only a small change in
[C/Fe] is seen as first dredge-up commences.

Our models are somewhat carbon rich compared to the observations. We
have chosen a primary that spends a long time on the TP-AGB and
therefore becomes very carbon-rich. Nevertheless, the model with
thermohaline mixing is a better representation of the data, which show
no evidence of a drop in [C/Fe] above $\log_{10}
L\mathrm{/L_{\odot}}=0.7$. The fact that there are no observed objects
with [C/Fe] greater than around 2.5 at luminosities less than
$\log_{10} L\mathrm{/L_{\odot}}=0.7$ suggests that accreted material
is mixed into the companion whilst it is still on the main sequence.

There is considerable scatter in the Lucatello et al. data. The
objects in this data set cover a spread of around two dex in
metallicity. In the framework of our model, we would expect there to
be scatter associated with the distribution of the parameters of the
binary systems (initial primary masses, mass ratios and separations).
This means there should be a spread in the abundances of the accreted
material. The dotted line in Fig.~\ref{fig:CFeversusL} shows the
results of a model with thermohaline mixing in which we decreased the
[C/Fe] ratio of the accreted material by 1 dex. It shows the same
trend as the original model but at lower [C/Fe], covering stars at the
lower end of the observed [C/Fe] range.

\begin{figure}
\centering
\includegraphics[width=8cm]{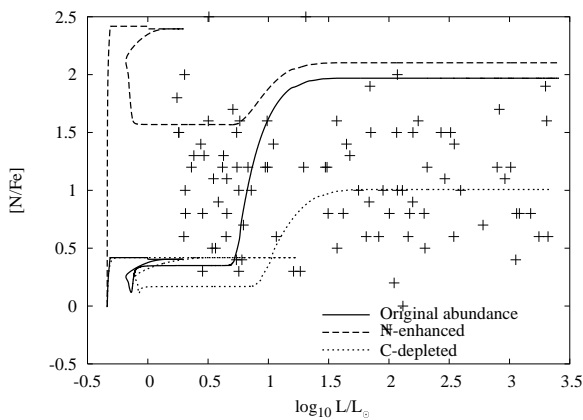}
\caption{Plot of the evolution of the [N/Fe] as a function of luminosity when thermohaline mixing is included. The solid line is for the model computed with the abundances predicted by the synthetic code. The dashed line is the model with the [N/Fe] ratio of the accreted material has been increased by 2 dex. The dotted line is the model with the [C/Fe] ratio of the accreted material decreased by 1 dex. The crosses represent the carbon-rich stars of the HERES sample \citep{2006ApJ...652L..37L}.}
\label{fig:NFeversusL}
\end{figure}

Figure~\ref{fig:NFeversusL} shows the evolution of the [N/Fe] ratio
for our original model with thermohaline mixing (solid line) against
the Lucatello et al. data. When first dredge-up occurs, we see a large
increase in the [N/Fe] ratio for the reasons outlined above. The size
of the increase at first dredge-up is dominated by the abundance of
carbon in the accreted material. This model covers the
upper end of the observed [N/Fe] values but only after dredge-up.  The
observations again show a large spread in [N/Fe] but no evidence of a
systematic increase at $\log_{10} L\mathrm{/L_{\odot}} \approx 0.7$.
The large number of stars with enhanced [N/Fe] at lower luminosities
in fact strongly suggests that the accreted material was already
nitrogen-rich. Figure~\ref{fig:NFeversusL} also shows two additional
tracks. In the model indicated with the dashed line we have boosted
the [N/Fe] ratio of the accreted material by 2 dex. This track passes
through the more nitrogen-rich points at low luminosity and reaches a
similar final [N/Fe] ratio to the original model. The second model
(dotted line) is the same as shown in Figure~\ref{fig:CFeversusL} in
which the [C/Fe] ratio of the accreted material is decreased by 1
dex. This model reaches a much lower final [N/Fe] ratio as should be
expected. Hence by varying the carbon and nitrogen abundances of the
accreted material the observed range of [N/Fe] values in the Lucatello
et al. sample can be covered reasonably well.  If thermohaline mixing
is not included, the [N/Fe] ratio shows a drop at $\log_{10}
L\mathrm{/L_{\odot}}\approx0.7$, similar to what is seen for [C/Fe] in
Fig.~\ref{fig:CFeversusL}, as the accreted material becomes mixed with
the convective envelope.

\section{Conclusion}
We have investigated the effects of thermohaline mixing on a low-metallicity, low-mass star that has accreted material from a carbon-rich companion. This material does not remain on the surface of the star as is commonly assumed. A thermohaline instability leads to the rapid mixing of this material with the rest of the star. We find that about 90\% of the star experiences mixing, while only 0.1\ms\ at the centre of the star remains unmixed. 

This has a very dramatic effect on the surface abundances of the star. The surface abundance of carbon is decreased by nearly an order of magnitude relative to the unmixed case. The models also display very different abundance patterns after the occurrence of first dredge-up. In the case of the model without thermohaline mixing, the post-first-dredge-up abundances are similar to those of the thermohaline mixing model when it has reached equilibrium after accretion. On the other hand, the model with thermohaline mixing displays a very large increase in its nitrogen abundance after first dredge-up.

Thermohaline mixing is an important component of models of CEMP stars. Without its inclusion, we would expect to see an abrupt change in the [C/Fe] ratio as a function of luminosity because of first dredge-up mixing the accreted material into the stellar interior. Such a change is not observed and thermohaline mixing is a possible explanation for this.
The observed [N/Fe] ratios can be reproduced by either model provided that there is a spread in the C and N abundances of the accreted material.

\begin{acknowledgements}
{ The authors thank the referee, Oscar Straniero, for his helpful comments}. RJS thanks Churchill College for his fellowship. EG and RGI thank the NWO for funding. 
\end{acknowledgements}
\bibliographystyle{aa}
\bibliography{6891}

\end{document}